\documentclass[12pt]{iopart}

\expandafter\let\csname equation*\endcsname\relax
\expandafter\let\csname endequation*\endcsname\relax
\usepackage{amsmath}
\usepackage{iopams}
\usepackage{graphicx}
\usepackage{citesort}

\begin{document}

\title{Widefield Microwave Imaging in Alkali Vapor Cells with sub-$100\,\mu$m Resolution}

\author{Andrew Horsley, Guan-Xiang Du, and Philipp Treutlein}

\address{Departement Physik, Universit\"{a}t Basel, Switzerland}

\ead{\mailto{andrew.horsley@unibas.ch}, \mailto{philipp.treutlein@unibas.ch}}


\begin{abstract}
We report on widefield microwave vector field imaging with sub-$100\,\mu$m resolution using a microfabricated alkali vapor cell.
The setup can additionally image dc magnetic fields, and can be configured to image microwave electric fields. Our camera-based widefield imaging system records 2D images with a $6\times 6$~mm$^2$ field of view at a rate of $10$~Hz. It provides up to $50\,\mu$m spatial resolution, and allows imaging of fields as close as $150\,\mu$m above structures, through the use of thin external cell walls. This is crucial in allowing us to take practical advantage of the high spatial resolution, as feature sizes in near-fields are on the order of the distance from their source, and represents an order of magnitude improvement in surface-feature resolution compared to previous vapor cell experiments. We present microwave and dc magnetic field images above a selection of devices, demonstrating a microwave sensitivity  of $1.4\,\mu\mathrm{T}\,\mathrm{Hz}^{-1/2}$ per $50\times50\times140\,\mu\mathrm{m}^3$ voxel, at present limited by the speed of our camera system.
Since we image $120\times120$ voxels in parallel, a single scanned sensor would require a sensitivity of at least $ 12\,\mathrm{nT}\,\mathrm{Hz}^{-1/2}$ to produce images with the same sensitivity. Our technique could prove transformative in the design, characterisation, and debugging of microwave devices, as there are currently no satisfactory established microwave imaging techniques. Moreover, it could find applications in medical imaging.
\end{abstract}

\vspace{2pc}
\noindent{\it Keywords}: microwave imaging, magnetic field imaging, atomic vapor cell

\maketitle

\section{\label{sec:introduction} Introduction}

Atomic vapor cells are one of the most versatile systems for measuring electromagnetic fields~\cite{Budker2002,Budker2007,Kitching2011}, and are at the heart of the most sensitive dc~\cite{Kominis2003,Dang2010} and rf~\cite{Savukov2005b,Savukov2014} magnetometers. Our group has recently developed a technique for imaging magnetic fields at microwave frequencies~\cite{Boehi2010a,Boehi2012,Horsley2013}, and alkali atoms in Rydberg states have been used for imaging microwave electric fields~\cite{Sedlacek2012,Sedlacek2013,Fan2014}. These techniques promise to have a transformative effect on the development, function and failure analysis of microwave devices in science and industry, as there is currently no established and satisfactory technique for imaging microwave fields. There is also significant interest in microwave sensing and imaging for medical applications, such as breast cancer screening~\cite{Fear2002,Nikolova2011,Chandra2015}.
However, while providing high field sensitivity, current vapor cell devices are limited to an exploitable spatial resolution on the millimeter scale. 

Here we report a new setup based on a $140\,\mu$m `ultrathin' vapor cell for high resolution imaging, providing $50\times50\times140\,\mu\mathrm{m}^3$ spatial resolution in the cell bulk, and allowing us to image fields as close as $150\,\mu$m above surfaces, thanks to a thin external wall. This represents an order of magnitude improvement in exploitable spatial resolution compared to previous vapor cell experiments, and allows us to enter the relevant regime for imaging fields of industrial microwave devices.
Our camera-based imaging technique allows us to record widefield 2D images at a rate of 10~Hz, which could be further improved to kHz rates using a faster camera system \cite{Gajdacz2013}. This allows us to record live movies of time-dependent processes, which would be rather difficult with a scanning probe system.
A particularly promising feature of our system is that it can be configured 
to also image microwave electric fields~\cite{Fan2014}.

Sub-millimeter spatial resolution has been reported in the vapor cell bulk for a number of sensing techniques~\cite{Xu2008,Zhao2008,Boehi2012,Horsley2013,Sedlacek2012,Sedlacek2013,Fan2014,Fescenko2014,Hakhumyan2011,Sargsyan2014}, but typical outer dimensions of cells have limited useable spatial resolution to the millimeter-scale or larger. Feature sizes in near-fields are on the order of the distance from the field source, meaning that, for example, micrometer-order spatial resolution cannot be exploited when performing sensing millimeters away from a field source. In order to resolve small structures on objects under investigation, it is crucial to measure fields at similarly small distances above the structures. There are many applications where sub-millimeter spatial resolution is essential, such as integrated microwave circuit characterisation~\cite{Wolff2006}, corrosion monitoring~\cite{Juzeliunas2006,Gallo2011,Gallo2012}, and in lab-on-a-chip environments for microfluidic analytical chemistry and bio-sensing~\cite{Xu2006a,Xu2006b,Xu2008,Harel2008}, and molecular imaging~\cite{Yao2010,Yu2012,Yao2013}.

\begin{figure}
\centering
\includegraphics[width=1\textwidth]{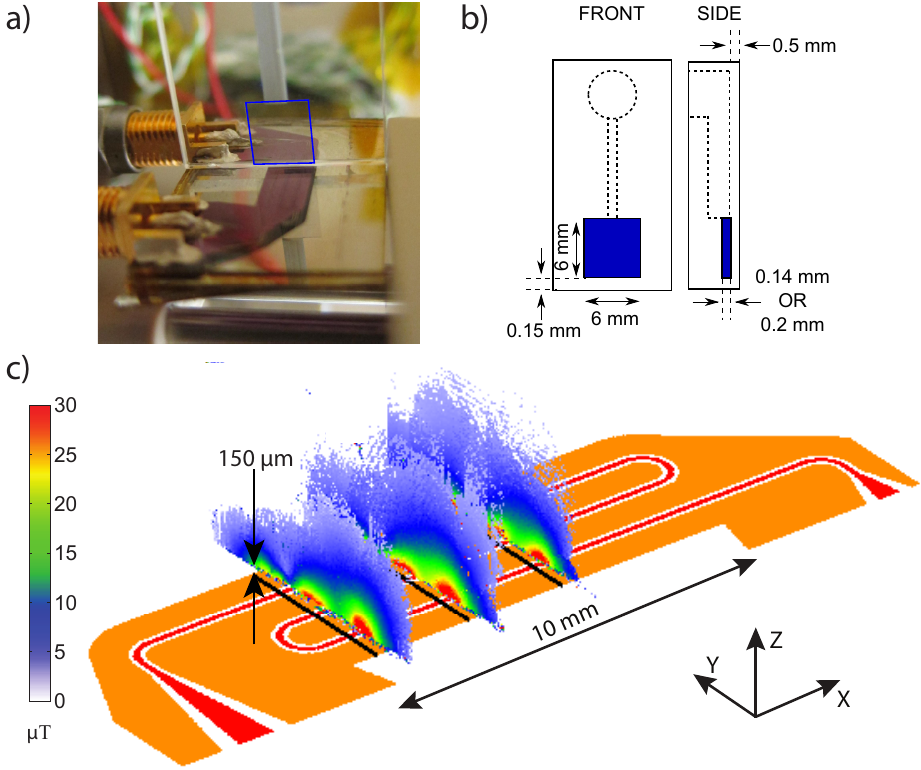}%
\caption{\label{fig:Zigzag3D}a) Photo of a microfabricated vapor cell positioned near a microwave test circuit. The cell chamber, highlighted in blue, allows us to record images with a $6\times 6$~mm$^2$ field of view.
b) Schematic of the vapor cells used in this work (not to scale). The cell chamber is shown in blue, and the etched  channel and through-hole are indicated with dotted lines. The key features are the extremely thin external cell walls: $500\,\mu$m on the side, and only $150\,\mu$m at the bottom end of the cell.
c) Experimentally obtained images of $|B_{\mathrm{mw}}|$, the absolute microwave magnetic field amplitude, in several cross-sections $150\,\mu$m above a coplanar waveguide arranged in a zigzag geometry (see Sec.~\ref{sec:zigzag}). The central signal line is shown in red, and the ground planes in orange. Black lines show the positions of the imaging planes on the chip. Differences in field shape at each position are due to differences in the relative phase of the microwave signal on the three loops of the signal line. The field at the middle imaging position is examined in more detail in Figure~\ref{fig:zigzag_field}.}
\end{figure}

We demonstrate our new high-resolution imaging system through the imaging of microwave magnetic near-fields above a selection of microwave circuits. As a demonstration of the flexibility of our setup, we also present vector-resolved images of the dc magnetic field above a wire loop.

\section{\label{sec:mw_reconstruction}Imaging Microwave Magnetic Fields in an Ultrathin Cell}

A photograph of our setup and typical microwave field images above a microwave integrated circuit are shown in figure~\ref{fig:Zigzag3D}. We use a microfabricated glass vapor cell with an inner thickness of $140~\mu$m to position a two-dimensional sheet of atomic Rubidium vapor near the microwave device under test (figure~\ref{fig:Zigzag3D}(a)). The cell features a $150~\mu$m thin side wall (figure~\ref{fig:Zigzag3D}(b)), which allows us to place the atoms at similarly small distance from the structure. The microwave field of the chip drives Rabi oscillations of frequency $\Omega_{\mathrm{Rabi}}(\vec{r})$ between hyperfine states of the atom, which depend on the projection of the local microwave field vector onto the direction of an applied uniform static magnetic field. The Rabi oscillations are recorded on a camera through the hyperfine state dependent absorption of a laser by the atomic vapor. Microwave field images obtained from the observed $\Omega_{\mathrm{Rabi}}(\vec{r})$ are shown in figure~\ref{fig:Zigzag3D}(c) in several cross sections of the chip.

A critical component in this work was the design of the two vapor cells used. The cells, identical in all but internal cell thickness, consist of two optically bonded $0.5$- and $1.5\times20\times90\,\mathrm{mm}$ pieces of Suprasil glass. As shown schematically in figure~\ref{fig:Zigzag3D}(b), a cell chamber with a thickness of either 140~$\mu$m or 200~$\mu$m is etched into one end of the 1.5~mm-thick piece. An etched channel connects the cell chamber to a through-hole, around which we attach a glass-to-metal transition with epoxy (Epotek-377). The key advance of our cells is their thin external walls (see figure~\ref{fig:Zigzag3D}(b)), as thin as $150\,\mu$m. In contrast to typical millimeter-scale vapour cell wall thicknesses, our thin walls allow us to image microwave near fields as close as $150\,\mu$m above microwave devices, and for the first time take practical advantage of our high spatial resolution.

We fill the cells with a 3:1 mixture of Kr and N$_2$ buffer gasses, with a typical filling pressure of 100~mbar measured at $T_{\mathrm{fill}}=22^{\circ}$C. The heavy Kr acts to localise the Rb atoms, improving our spatial resolution and limiting depolarising Rb collisions with the cell walls~\cite{Vanier1989}. The N$_2$ gas is included for quenching effects~\cite{Chevrollier2012,Rosenberry2007}.

\begin{figure}
\centering
\includegraphics[width=0.8\textwidth]{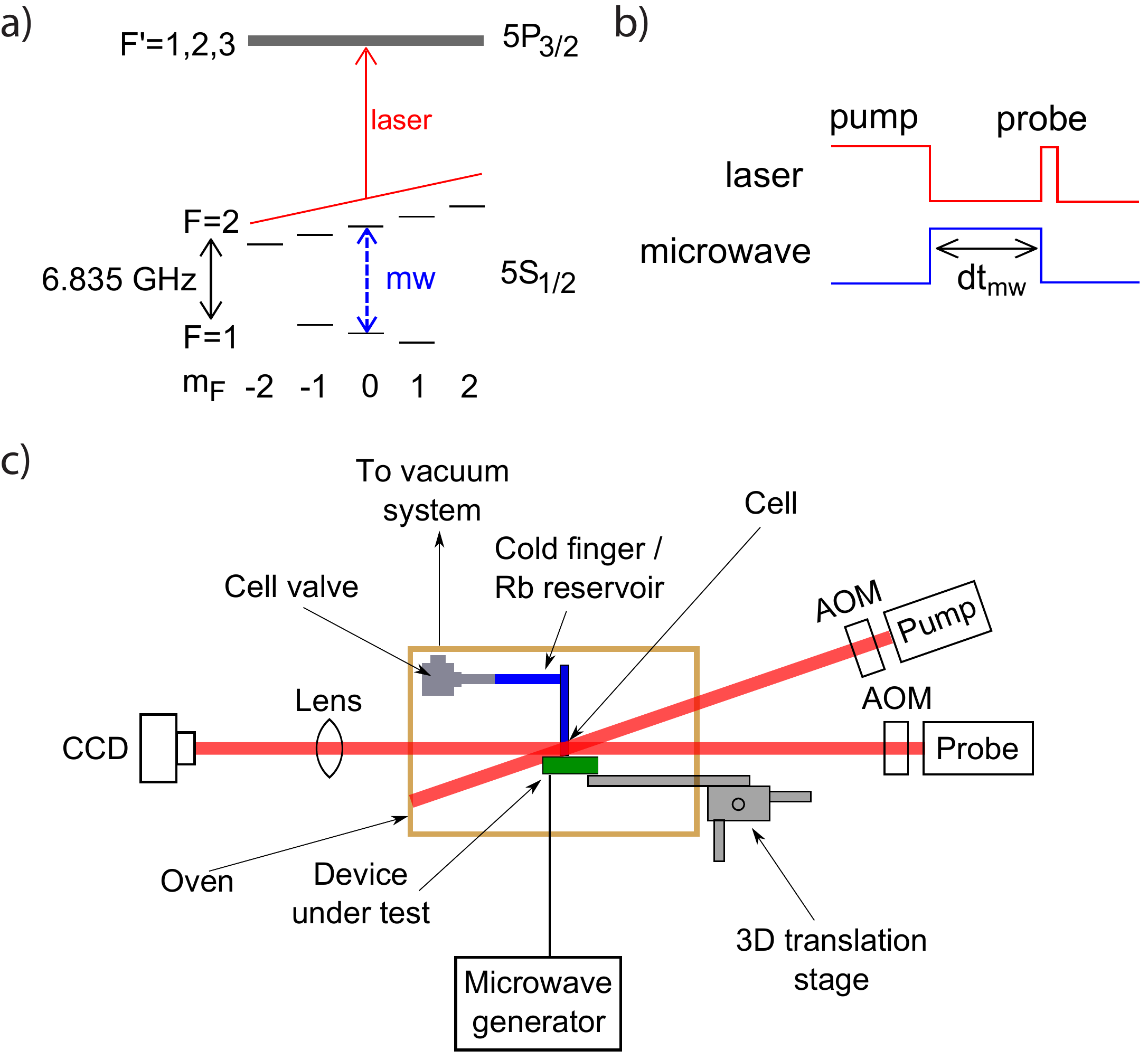}%
\caption{\label{fig:setup_highres}a) The $^{87}$Rb D$_2$ line. Due to Doppler and collisional broadening on the optical transitions, the $F'$ excited state levels are not resolved. Transitions between the Zeeman-split m$_F$ levels of the ground state hyperfine structure can be individually addressed by tuning the microwave frequency. We use Rabi oscillations driven on the `clock' transition to detect the microwave magnetic field; b) The experiment sequence; c) The experimental setup. AOM = acousto-optical modulator.}
\end{figure}

A schematic of our cell is shown in Figure~\ref{fig:setup_highres}(c). The cell and microwave device are placed inside an oven, with operating temperatures of $130^{\circ}$C to $140^{\circ}$C chosen to give an optical depth of $OD\approx1$~\cite{Siddons2008,Weller2011,Zentile2014}. The Rb vapor density is controlled by a cold finger wrapped around the end of the glass-to-metal transition, and the $10^{\circ}$C temperature gradient between the cold finger and the cell helps reduce the deposition rate of Rb and other contaminants on the cell windows. The experiment is surrounded by a cage of Helmholtz coils, which cancel the Earth's magnetic field, and provide a static field of 1-2~G along the $X$, $Y$, or $Z$ axes. This field serves as the quantisation axis, and the resulting $\sim$MHz Zeeman splitting of the $^{87}$Rb hyperfine ground state transitions allows each transition to be individually addressed by tuning the microwave frequency.

Figure~\ref{fig:setup_highres}(a) shows the hyperfine structure of the D$_2$ line of $^{87}$Rb and the relevant microwave and optical transitions involved.
We produce images of each of the polarisation components of the microwave magnetic field, using Rabi oscillations driven on the $|F=1,m_F=0\rangle \rightarrow |F=2,m_F=0\rangle$ `clock' transition of the $^{87}$Rb hyperfine ground state~\cite{Boehi2010a,Boehi2012,Horsley2013}. The Rabi frequency on this transition is given by $\Omega_{\mathrm{Rabi}}(\vec{r})=\frac{\mu_B}{\hbar}B_{\mathrm{mw}}(\vec{r})$, where $B_{\mathrm{mw}}(\vec{r})= (\vec{B}_{\mathrm{mw}}(\vec{r})\cdot \vec{B})/|\vec{B}|$ is the projection of the local microwave field vector $\vec{B}_{\mathrm{mw}}(\vec{r})$ onto the direction of an applied uniform static magnetic field $\vec{B}$.
Imaging with the static field pointing in the $X$, $Y$, and $Z$ directions allows us to image each polarisation component in turn. Using atoms as sensors, our technique avoids the significant calibration problem in other microwave sensors~\cite{Zelder2008}, relating the field to a measured oscillation frequency and well-known fundamental physical constants. Data taking is fast, due to the parallel nature of the measurement (imaging as opposed to scanning). By applying an external static magnetic field, we have imaged microwave fields from 2.3~GHz to 26.4~GHz with a single device~\cite{Horsley2015a}. The technique is applicable to microwave devices of all types, recently showing success in characterising and debugging microwave cavities in high-performance miniaturised atomic clocks~\cite{Horsley2013,Ivanov2014,Affolderbach2015}.

\begin{figure}
\centering
\includegraphics[width=0.8\textwidth]{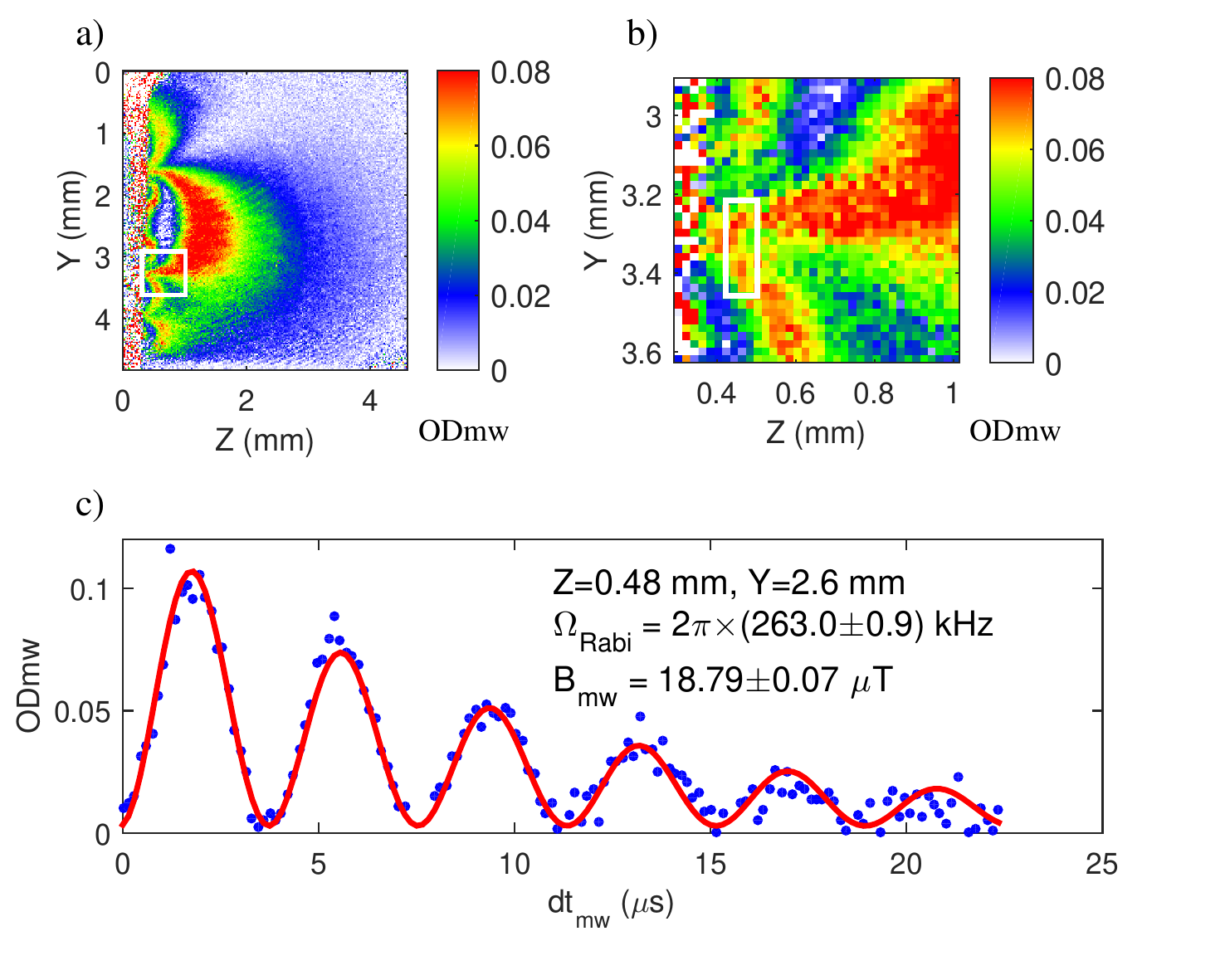}%
\caption{\label{fig:OD_P2}a) Example image, and b) zoomed in section, of OD$_{\mathrm{mw}}$, the change in optical density produced by a microwave pulse, in this case $dt_{\mathrm{mw}}=4.65\,\mu$s long, from a microwave device located to the left of the image. The zoomed-in section is indicated by the white box in (a). OD$_{\mathrm{mw}}$ images show contour lines of the microwave magnetic field. The smallest feature size, highlighted in the zoomed-in image, is only $60-80\,\mu$m peak-to-trough; c) OD$_{\mathrm{mw}}$ at $Z=0.48$~mm, $Y=2.6$~mm, showing Rabi oscillations as the microwave pulse length is scanned. }
\end{figure}

We begin an experiment sequence, shown in figure~\ref{fig:setup_highres}(b), by preparing the atoms in the $^{87}$Rb $F=1$ ground state with a 1~ms optical pumping pulse. Through frequent ($\sim10^9\,\mathrm{s}^{-1}$) collisions with the buffer gas, Rb atoms sample the entire velocity space over the course of the optical pumping pulse (and also the subsequent probe pulse). We typically see a 30$\%$ reduction in OD due to optical pumping. The optical pumping efficiency is below 100$\%$ due to several factors: radiation trapping; collisional broadening of the optical line; absorption due to $^{85}$Rb; and the detuning of the lasers from the collisionally shifted $^{87}$Rb and $^{85}$Rb optical lines~\cite{Horsley2015thesis}. We drive Rabi oscillations by injecting a microwave pulse of length $dt_{\mathrm{mw}}$ into the microwave device under test. We then image the resulting repopulation of the $F=2$ state with a $dt_{\mathrm{probe}}=0.3\,\mu$s probe pulse using absorption imaging, which selectively detects the $F=2$ state~\cite{Horsley2013,Horsley2013a}. The optical pumping and probing is performed with two separate 780~nm diode lasers, frequency stabilised to the $F=2 \rightarrow F'=2,3$ crossover peak of the $^{87}$Rb D$_2$ line, red-shifted by an AOM $80\,\mathrm{MHz}$ from the stabilisation point, and with intensities of $120\,\mathrm{mW}/\mathrm{cm}^2$ and $30\,\mathrm{mW}/\mathrm{cm}^2$, respectively. The short probe pulse length ensures that optical pumping due to the probe pulse is minimal. We take reference images to account for short and long term drifts, and combine the images to give an image of OD$_{\mathrm{mw}}$, the change in optical density (OD) induced by the microwave pulse. An example OD$_{\mathrm{mw}}$ image is shown in figure~\ref{fig:OD_P2}(a). The inhomogeneous microwave field drives Rabi oscillations at different rates across the image, which form patterns in OD$_{\mathrm{mw}}$ following the contour lines of the microwave field. Atoms along the outermost (mostly) red line of figure~\ref{fig:OD_P2}(a) are at the peak of their first Rabi oscillation, corresponding to maximal repopulation of the absorptive $F=2$ state. The inner red line corresponds to a region of higher field, where atoms are at the peak of their second oscillation. We take multiple OD$_{\mathrm{mw}}$ images, scanning $dt_{\mathrm{mw}}$, to produce OD$_{\mathrm{mw}}$ movies. A sample of these movies are available online, with the frame rate matching the 10~Hz image acquisition rate of our experiment. The counter on top of the movies indicates the microwave pulse duration.
As shown in figure~\ref{fig:OD_P2}(c), each pixel in these movies has an oscillating signal which we can fit to obtain the local microwave field strength.

\section{\label{sec:hellma_cells}Spatial Resolution}

The longitudinal spatial resolution of our imaging setup is set by the 140- or $200\,\mu$m thickness of the cell. The buffer gas pressures set a similar transverse spatial resolution, by determining the distance an atom can diffuse over the course of a measurement~\cite{Vanier1989}. At $T=135^{\circ}$C, we estimate the r.m.s diffusion distance during a $dt_{\mathrm{mw}}=T_1=8.8\,\mu$s measurement to be $\Delta x = \sqrt{2 D dt_{\mathrm{mw}}} = 50\,\mu$m. Here, $T_1$ is the hyperfine population relaxation time constant in the cell. The diffusion coefficient, $D$, is given by $1/D=1/D_{\mathrm{Kr}}+1/D_{\mathrm{N_2}}$, where $D_{\mathrm{Kr}(\mathrm{N_2})}=D_{0,\,\mathrm{Kr}(\mathrm{N_2})} \frac{P_0}{P_{\mathrm{Kr}(\mathrm{N_2})}} (\frac{T}{T_0})^{3/2}$. $P_0=1$~atm, $P_{\mathrm{Kr}(\mathrm{N_2})}$ is the Kr (N$_2$) pressure, and we have used $D_{0,\,Kr}=0.068\,\mathrm{cm}^2/\mathrm{s}$~\cite{Chrapkiewicz2014} and $D_{0,\,N_2}=0.159\,\mathrm{cm}^2/\mathrm{s}$~\cite{Ishikawa2000}. These are order-of-magnitude increases in spatial resolution compared to previous imaging experiments~\cite{Boehi2012,Fan2014,Mikhailov2009,Fescenko2014}.

Peak-to-trough feature sizes as small as $70\pm10\,\mu$m can be seen in the OD$_{mw}$ images, approaching the estimated diffusion-limited spatial resolution. An example is shown in figure~\ref{fig:OD_P2}.

\section{Microwave Field Sensitivity}\label{sec:sensitivity}

The $6\times6$~mm cell can be thought of as an array of $N_{\mathrm{sens}}=120\times120$ sensors, with each sensor corresponding to a $50\,\mu\mathrm{m}\times 50\,\mu\mathrm{m}\times 140\,\mu\mathrm{m}$ voxel. The sensor size is given by the diffusion-limited spatial resolution, with a sensor volume $V=1.8\times10^{-7}\,\mathrm{cm}^3$. To estimate our experimental sensitivity, we examined CCD pixels binned in $2\times2$ blocks, corresponding to an area of $42\times42\,\mu\mathrm{m}^2$, slightly smaller than the sensor size. The fitting error to our microwave Rabi data was as low as $21\,$nT per $2\times2$ pixels, giving an estimated sensitivity of $\delta B_{\mathrm{mw}}^{\mathrm{exp}}= 1.4\,\mu\mathrm{T}\,\mathrm{Hz}^{-1/2}$ per sensor, taking into account the 4440~s measurement time (148 averaged runs). Integrating over a larger volume would give an increase in sensitivity, at the expense of spatial resolution.

We record data for all of the sensors in our array simultaneously. Compared to creating an image by scanning a single sensor, this improves our data taking speed, by a factor of at least $N_{\mathrm{sens}}$, or four orders of magnitude. The effective sensitivity is therefore significantly improved by our parallel imaging, and a single, scanned sensor would require a sensitivity of at least $\delta B_{\mathrm{mw}}^{\mathrm{exp}}/\sqrt{N_{\mathrm{sens}}}= 12\,\mathrm{nT}\,\mathrm{Hz}^{-1/2}$ to produce an image with the same sensitivity. Parallel imaging is also more suitable than scanning for applications requiring high temporal resolution over an image.

We can compare our experimental sensitivity with the photon shot noise limited sensitivity. Assuming $\Omega_{\mathrm{Rabi}} \,dt_{\mathrm{mw}}\ll \pi$, we have \cite{Horsley2015thesis}
\begin{equation}
\delta B_{\mathrm{photon}}= \sqrt{\frac{dt_{\mathrm{run}}}{N_{\mathrm{shots}}}}\frac{2}{dt_{\mathrm{mw}}}\frac{\hbar}{\mu_B} \frac{OD_{\mathrm{min}}}{OD_{\mathrm{mw}}^{\mathrm{max}}}\,\exp(dt_{\mathrm{mw}}/\tau_2).
\end{equation}
We first calculate $\delta B_{\mathrm{photon}}$ for conditions matching our experiment parameters: an experiment run of $N_{\mathrm{shots}}=150$ shots taking a time $dt_{\mathrm{run}}=30\,\mathrm{s}$; $dt_{\mathrm{mw}}=22.5\,\mu$s; an atomic coherence lifetime $\tau_2=7.8\,\mu$s; a measured operating temperature of $T_{\mathrm{res}}=140^{\circ}$C; total buffer gas pressure of  $P_{\mathrm{fill}}=100\,\mathrm{mbar}$; optical pumping resulting in $1/3$ of the atomic population residing in each of the $F=1$ ground states, such that OD$_{\mathrm{mw}}^{\mathrm{max}}=\tfrac{1}{3}OD_{87}=0.24$, where $OD_{87}$ is the OD of the $^{87}$Rb in the cell; and a photon shot noise limited $OD_{\mathrm{min}}=\sqrt{2}[\,Q\, I_{\mathrm{probe}}\,e^{-OD}\, A\, dt_{\mathrm{probe}}/(\hbar \omega_L)]^{-1/2}=1.0\times10^{-2}$, where $\omega_L$ is the laser frequency, $Q=0.27$ is the camera quantum efficiency, $I_{\mathrm{probe}}= 30\,\mathrm{mW}/\mathrm{cm}^2$ is the probe intensity, $dt_{\mathrm{probe}}=0.3\,\mu$s is the probe duration, and the $2\times2$ pixel area is $A=42\times42\,\mu\mathrm{m}^2$. This gives us $\delta B_{\mathrm{photon}}=0.45\,\mu\mathrm{T}\,\mathrm{Hz}^{-1/2}$. The exact operating temperature was unclear, however, with measurements of the OD indicating that the operating temperature may have been closer to $T_{\mathrm{res}}=130^{\circ}$C, which would give $\delta B_{\mathrm{photon}}=0.28\,\mu\mathrm{T}\,\mathrm{Hz}^{-1/2}$. We therefore conclude that our measured $\delta B_{\mathrm{mw}}^{\mathrm{exp}}= 1.4\,\mu\mathrm{T}\,\mathrm{Hz}^{-1/2}$ is $3-5$ times the photon shot noise limit determined by our experiment parameters. Analysis of OD$_{\mathrm{mw}}$ noise in the absence of a microwave field indicates that half of the $\delta B_{\mathrm{mw}}^{\mathrm{exp}}$ in excess of $\delta B_{\mathrm{photon}}$ is caused by imaging noise, due to factors such as camera readout noise, and fluctuations in the intensities and frequencies of the lasers. Sources for the second half of the excess noise include fitting errors and timing jitter in the experiment sequence. We also note that we perform the imaging without magnetic shielding.

The optimal photon shot noise limited sensitivity, $\delta B_{\mathrm{photon}}^{\mathrm{opt}}= 0.08\,\mu\mathrm{T}\,\mathrm{Hz}^{-1/2}$, is reached for $T_{\mathrm{res}}=130^{\circ}$C, $P_{\mathrm{fill}}=60\,\mathrm{mbar}$, and with the laser tuned to the buffer-gas-shifted $^{87}$Rb $F=2\rightarrow F'=2$ line. Assuming that we can reach the photon shot noise limit, by reducing the excess noise from the above sources, we could expect a factor of $17.5$ improvement in sensitivity with only minor modifications to our setup.

An improvement in sensitivity of several orders of magnitude is possible with more involved modifications. We are operating $5\times10^5$ above the atomic projection noise limit, the ultimate sensitivity limit of an atom-based sensor~\cite{Budker2007}. Both $\delta B_{\mathrm{mw}}^{\mathrm{exp}}$ and $\delta B_{\mathrm{photon}}$ are limited by the camera readout speed and data saving time, which give a poor experiment duty cycle (10 OD$_{\mathrm{mw}}$ images per second) and result in the atoms sitting uninterrogated for the vast majority of the time. This could be dramatically sped up with a different camera and camera operation mode, and we note that $50\times50$ pixel imaging of ultracold atoms has been reported with a continuous frame rate of $2500$~fps~\cite{Gajdacz2013}. Approaching the atomic projection noise limit will ultimately require moving to a quasi-continuous measurement scheme, likely based on Faraday rotation~\cite{Allred2002,Gajdacz2013}, and perhaps replacing the CCD camera with an array of photodiodes.

\section{\label{sec:imaging} Imaging Microwave Fields Above Test Structures}

In order to characterise and demonstrate our imaging system, we created three demonstration structures. The structures, shown in figures~\ref{fig:cpw_comparison}-\ref{fig:zigzag_field}, respectively, are: a coplanar waveguide (CPW); a waveguide making several bends across its substrate, which we dubbed the `Zigzag' chip; and a split-ring resonator (SRR). All of the microwave field measurements were made using the $140\,\mu$m cell.

For imaging, the chip is generally placed perpendicular to the end of the vapor cell, as shown in figure~\ref{fig:setup_highres}(b). For chips built on a transparent  or reflective substrate, operation in a second mode is also possible, with the chip placed in front of and parallel to the vapor cell, as shown in figure~\ref{fig:SRR_field}(a).

We use the program Sonnet to perform a simulation of the microwave propagation on our structures using the Method of Moments. This technique is well suited for our mostly planar structures, excited at a single frequency. The program outputs the current distribution on the chip, from which we compute the magnetic near-fields using the Biot-Savart law. The only free parameters in comparisons with measurement were the amplitude of the input microwave signal and the exact position of the cell relative to the chip.

\subsection{\label{sec:CPW} The Coplanar Waveguide}

\begin{figure}
\centering
\includegraphics[width=0.8\textwidth]{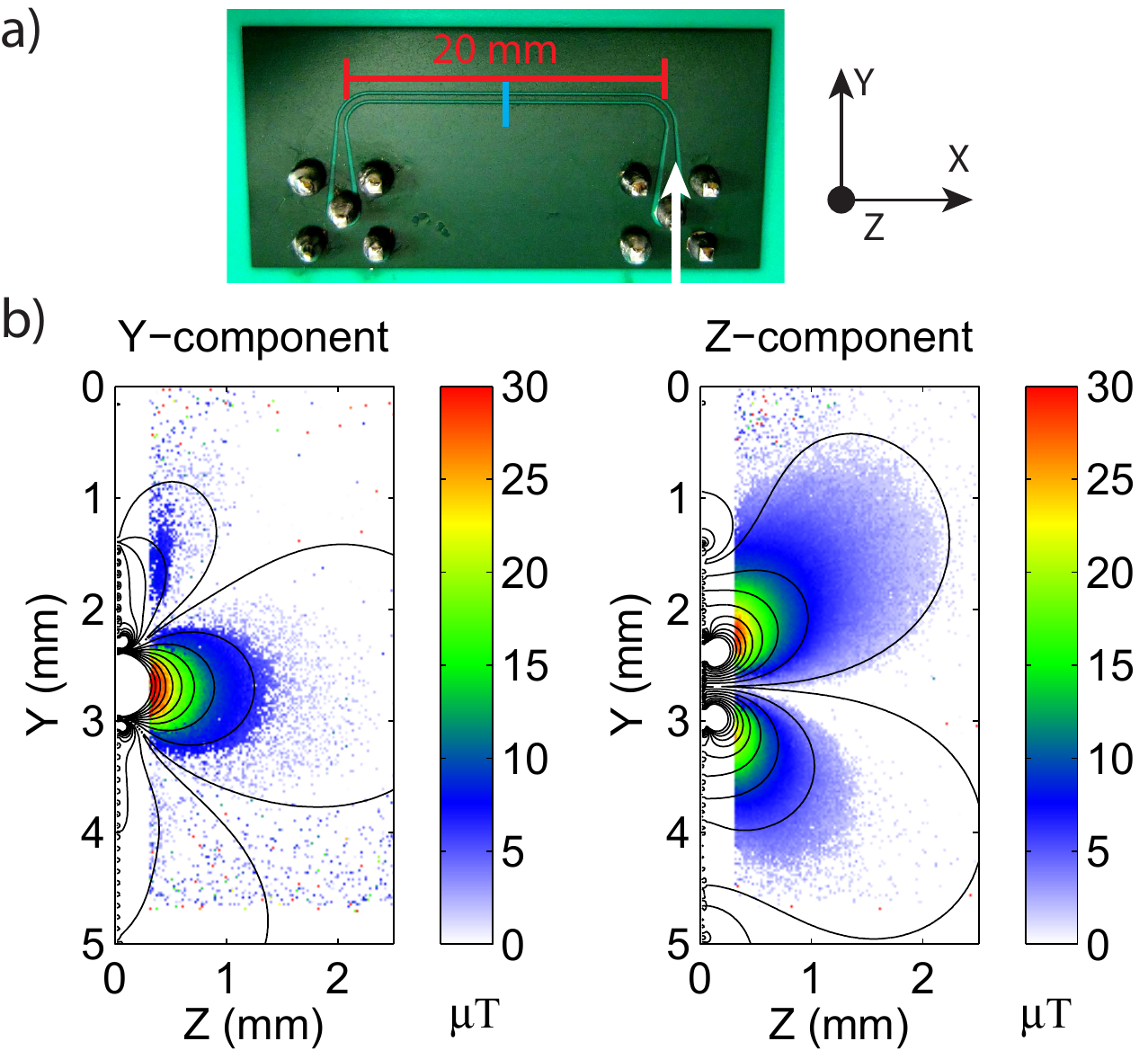}%
\caption{\label{fig:cpw_comparison}a) Photo of the CPW chip, with the orientation of the chip in relation to the coordinate system defined by the imaging cell shown on the right. The approximate position of the imaging plane is indicated by a blue line, and a white arrow indicates the microwave insertion port; b) Experimentally obtained images of the $Y$ and $Z$ components of the microwave magnetic field above the CPW. The waveguide surface is at approximately $Z=0$. The simulated microwave field is shown in black contour lines, starting at 1~$\mu$T for the outermost line and increasing in 5~$\mu$T steps inwards.}
\end{figure}

CPWs are a ubiquitous building block of microwave circuits~\cite{Wolff2006}, and provide a simple structure which can be readily and robustly compared with simulations. The CPW used in this work, shown in figure~\ref{fig:cpw_comparison}(a), has a 500~$\mu$m wide central signal strip, with 105~$\mu$m gaps to ground planes on either side. Figure~\ref{fig:cpw_comparison}(b) shows images of the $Z$ and $Y$ components of the CPW microwave magnetic field (the very weak $X$ component was not imaged). Simulations of the microwave field are shown as overlaid contour lines. The slight asymmetry is related to the bends in the wires. The good agreement with the simulated field demonstrates the reliability of the imaging technique. Discrepancies may be due to imperfect coupling into the waveguide, and the use of a finite mesh size for modelling the microwave field through the bends. The images in Figure~\ref{fig:cpw_comparison}.b demonstrate the importance of thin external vapor cell walls: a vapor cell with standard millimeter-scale external walls would see none of the interesting features.

\subsection{\label{sec:zigzag} The Zigzag Chip}

\begin{figure}
\centering
\includegraphics[width=1\textwidth]{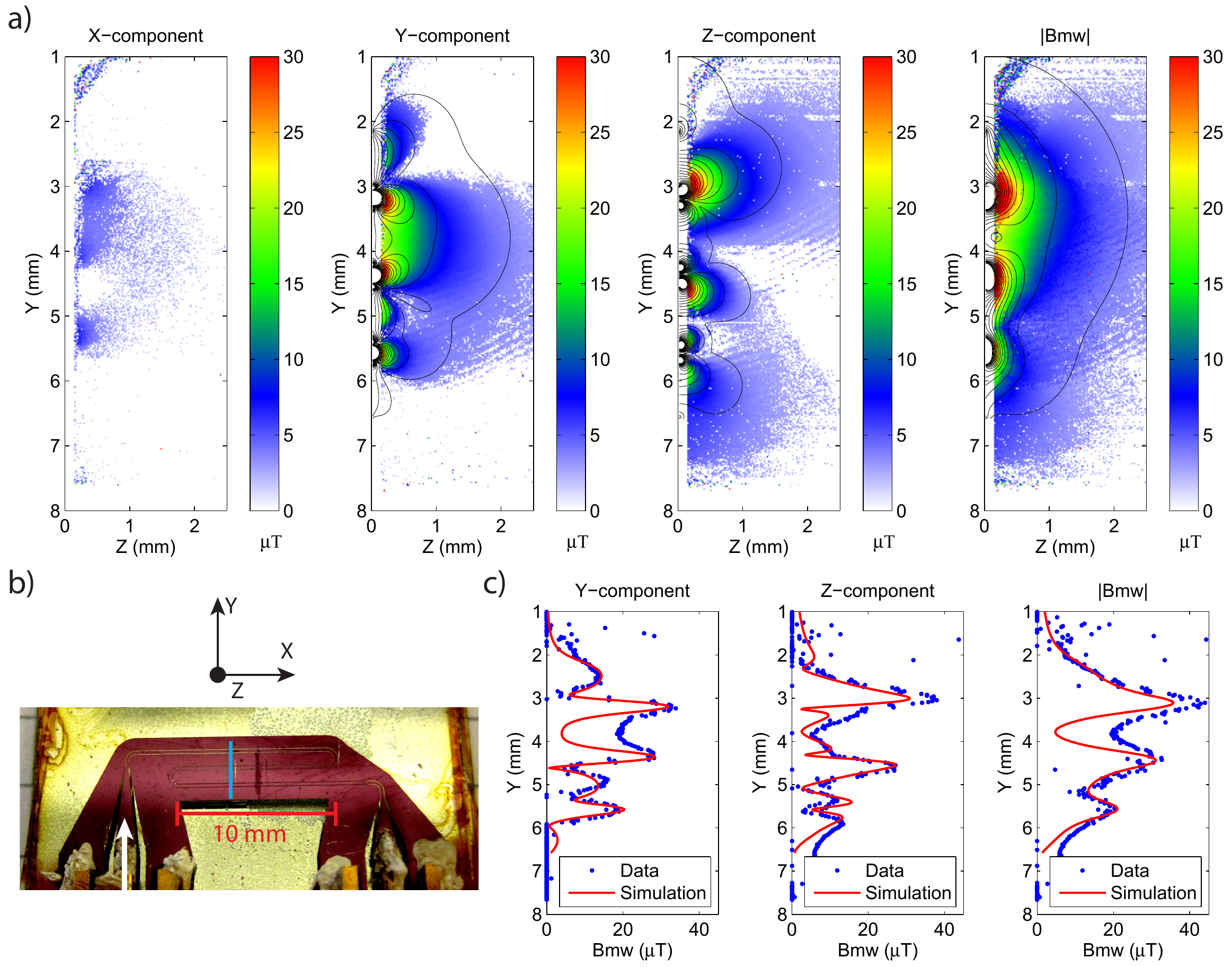}%
\caption{\label{fig:zigzag_field} a) Experimentally obtained images of the $X$, $Y$, and $Z$ components of the microwave magnetic field above the Zigzag chip. The magnitude of the microwave field, $\lvert B_{\mathrm{mw}} \rvert=\sqrt{B_X^2+B_Y^2+B_Z^2}$, is also shown on the far right. The waveguide surface is at approximately $Z=0$. The simulated microwave field is shown in black contour lines, starting at 2~$\mu$T for the outermost line and increasing in 3~$\mu$T steps inwards; b) Photo of the Zigzag chip. The approximate position of the imaging plane is indicated by a blue line, and a white arrow indicates the microwave insertion port; c) Cross-sections of the experimentally obtained microwave field (blue dots) approximately $250\,\mu$m above the Zigzag chip surface, and comparison to simulation (red lines).}
\end{figure}

The Zigzag chip, shown in figure~\ref{fig:zigzag_field}(b), has smaller and more complex features than the CPW, allowing us to highlight the spatial resolution of our setup. The Zigzag waveguide has a 200~$\mu$m thick central signal strip, with 50~$\mu$m gaps to ground planes either side. The waveguide goes through two bends, resulting in a cross-section in the imaging plane containing three waveguide sections, each separated by 900~$\mu$m. Figure~\ref{fig:Zigzag3D} shows quasi-2D slices of the absolute microwave amplitude, $|B_{\mathrm{mw}}|$, at three positions above the Zigzag chip. The variation in field shape between the positions is due to the standing wave produced in the waveguide. Figure~\ref{fig:zigzag_field}(a) then examines the middle imaging plane of figure~\ref{fig:Zigzag3D} (indicated by the blue line in figure~\ref{fig:zigzag_field}(b)) in more detail, showing images of each of the polarisation components of the microwave field above the chip, which are compared with contour lines from simulation. Cross-sections of the field near the edge of the vapor cell are shown in Figure~\ref{fig:zigzag_field}.c. The wide field of view in figures~\ref{fig:Zigzag3D} and~\ref{fig:zigzag_field} ($>6\,\mathrm{mm}$) was obtained by stitching two sets of images together.


There is general agreement between the measured and simulated fields in figure~\ref{fig:zigzag_field}, but not for all features. The amplitude of the simulated $X$ component of the field is well below the experimental sensitivity, and the measured $X$ component of the field is likely to be some projection of the $Y$ and $Z$ components, caused by imperfect orthogonality between the chip, cell, and coil axes. Additionally, as seen in the cross-sections in figure~\ref{fig:zigzag_field}(c), the measured microwave field is much broader than the simulation around $Y=3\,\mathrm{mm}$ to $Y=4.5\,\mathrm{mm}$. Given the spatial resolution shown at $Y=5.6\,\mathrm{mm}$, it is reasonable to conclude that this broadening is a real feature of the microwave field. It is unlikely to be due to perturbations induced by the vapor cell, for which we were unable to measure any effect with the Zigzag or CPW chips. Such discrepancies highlight the difficulty of accurately manufacturing and simulating even relatively simple structures such as the Zigzag chip, and the need for direct measurements.

\subsection{\label{sec:SRR} The Split-Ring Resonator}

\begin{figure}[t]
\centering
\includegraphics[width=1\textwidth]{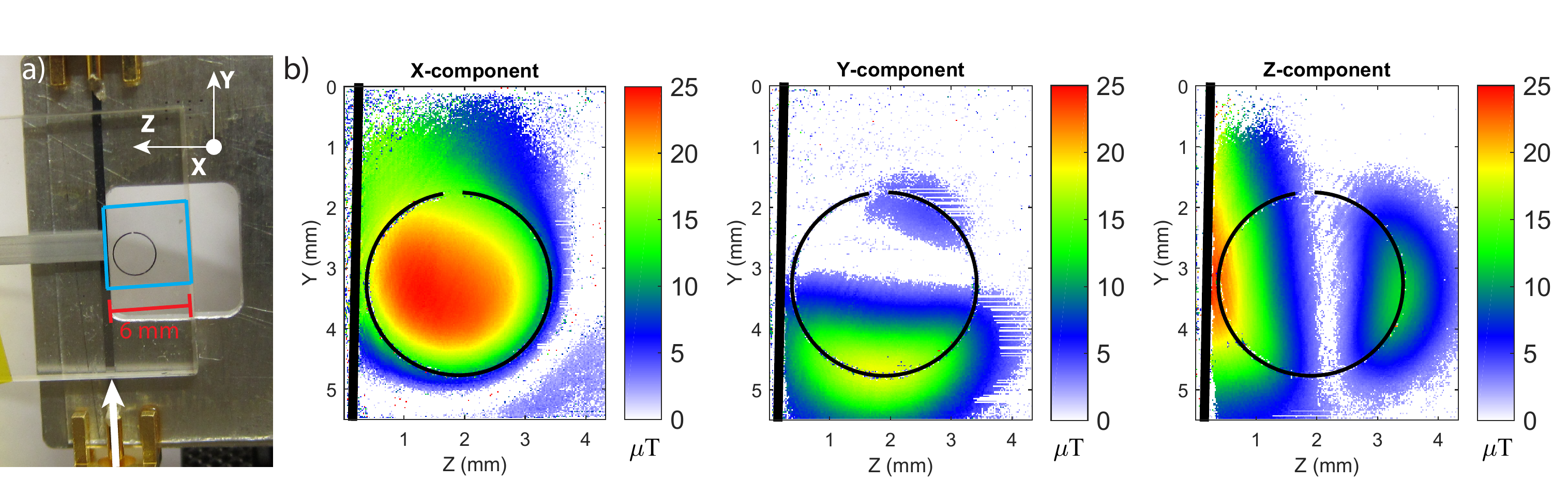}%
\caption{\label{fig:SRR_field} a) Photo of the SRR chip, demonstrating a second operation mode of the imaging setup, with the glass cell parallel to the transparent chip surface; b) Experimentally obtained images of the $X$, $Y$, and $Z$ components of the microwave magnetic field above the split-ring resonator (SRR). The waveguide surface is parallel to, and a few millimeters in front of, the cell. Black outlines show the positions of the signal line and ring.}
\end{figure}

The SRR chip, shown in figure~\ref{fig:SRR_field}(a), consists of a signal line coupling inductively into a split ring. The split-ring is built on a transparent glass substrate, allowing us to operate in a second mode, with the SRR placed in front of and parallel to the vapor cell. The resonator linewidth was $160\pm20$~MHz, corresponding to a quality factor of $40\pm5$.

The presence of the vapor cell significantly changed the properties of the SRR, by filling the space around the resonator with a glass dielectric. We used this to tune the resonance frequency to match the 6.835~GHz splitting of the $^{87}$Rb ground states, adjusting the gap between the cell and the SRR until the resonance was in the desired position. A shift of $1\,\mu$m corresponded to a shift in resonance of 5.7~MHz. Note that we were unable to detect any influence of the cell on the CPW or Zigzag chips.

The SRR field is shown in figure~\ref{fig:SRR_field}(b). Like in a solenoid, the SRR field is strongest inside the split-ring, parallel to the split-ring axis in the $X$ direction. The field then turns outward, seen in the $Y$ and $Z$ component images, before returning with a less-dense flux in the $X$ direction outside the split-ring. The minima in the centres of the $Y$ and $Z$ components are because the field lines travel out from the field centre, and so cancel out along the central axes. The lopsided nature of the $Y$ component is due to the presence of the split in the ring.

\section{\label{sec:dc_imaging} Vector Imaging Of A DC Magnetic Field}

\begin{figure}[ht]
\centering
\includegraphics[width=1\textwidth]{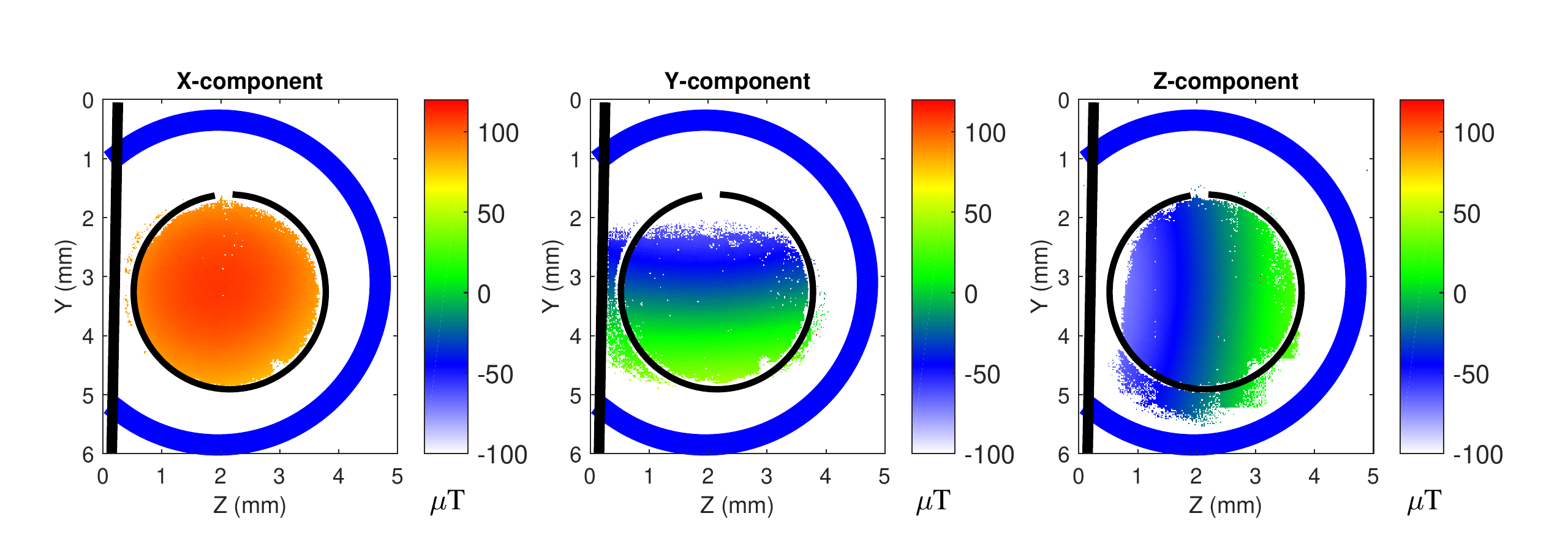}%
\caption{\label{fig:dc_field} Experimentally obtained images of the $X$, $Y$, and $Z$ components of a dc magnetic field X~mm above a wire loop. Positive and negative field values represent opposite directions. The field of view corresponds to the $X$ component of the SRR microwave magnetic field, which was used to drive the Ramsey oscillations used to image the dc field. Outlines show the positions of the current loop (blue) and SRR (black). The coordinate system is the same as shown in figure~\ref{fig:SRR_field}(a).}
\end{figure}

Our imaging technique can be adapted to measure dc magnetic fields. We use a Ramsey sequence~\cite{Horsley2013}, where the single microwave pulse of the above Rabi sequence is replaced by two $\pi/2$ pulses separated by a time $dt_{\mathrm{Ramsey}}$. Driving oscillations on the magnetic field sensitive $|F=1,m_F=1\rangle \rightarrow |F=2,m_F=1,2\rangle$ transitions, the oscillation frequency of the Ramsey fringes is equal to the detuning of the microwave from resonance, allowing us to measure the Zeeman shift induced by the applied dc magnetic fields. We can then use the Breit-Rabi formula to obtain the dc field of interest.

To detect individual vector components of a field of interest $\vec{B}$, we apply a second dc magnetic field of strength $C \gg B$. In this way, we are primarily sensitive to the component of $\vec{B}$ that is parallel to $\vec{C}$. For $\vec{C}$ along the $X$ axis, the measured field, $B_{\mathrm{meas}}$, is~\cite{Budker2007}
\begin{equation}
B_{\mathrm{meas}} = \sqrt{(C + B_X)^2 +  B_Y^2 +  B_Z^2} \approx C + B_X.
\end{equation}
We can obtain $C$ in a separate reference measurement, and subtract this from $B_{\mathrm{meas}}$ to obtain $B_X$. The full vector magnetic field can be obtained by imaging with the $C$ field applied along each of the $X$, $Y$, and $Z$ axes.

Figure~\ref{fig:dc_field} shows images of the dc field above a 2~mm diameter wire loop, taken using the $200\,\mu$m-thick cell. Again, we see a solenoid-like field, with a strong, uniform $X$ component, and the field turning outwards in the $Y$ and $Z$ components. Following the discussion on microwave sensitivity in section~\ref{sec:sensitivity}, fitting uncertainties give a sensitivity as small as $\delta B_{\mathrm{dc}}^{\mathrm{exp}} = 1.6\,\mu\mathrm{T}\,\mathrm{Hz}^{-1/2}$ for a $40\times40\times200\,\mu$m sensor. 
As discussed in section~\ref{sec:sensitivity}, the dominant limiting factor is our poor experiment duty cycle, improving which promises orders of magnitude increases in sensitivity.

\section{\label{sec:outlook} Conclusions and Outlook}

We have demonstrated a new setup for high resolution imaging of electromagnetic near fields, through the imaging of microwave fields above a variety of microwave devices, and the dc magnetic field above a wire loop. Microwave imaging is performed with a $120\times120$ array of $50\times50\times140\,\mu\mathrm{m}^3$ sensors, with the sensor size given by atomic diffusion during a measurement and the $140\,\mu\mathrm{m}$ cell thickness. The sensitivity per sensor, $\delta B_{\mathrm{mw}}^{\mathrm{exp}} =  1.4\,\mu\mathrm{T}\,\mathrm{Hz}^{-1/2}$, is primarily limited by the experiment duty cycle, and improvements of several orders of magnitude should be achievable. We obtained a similar sensitivity for dc magnetic field imaging in a $200\,\mu$m thick cell. The setup allows us to image fields as close as $150\,\mu$m above surfaces, resulting in an order of magnitude increase in the resolution of surface features compared to previous vapor cell sensors. To our knowledge, this is the first vapor cell with such thin walls, and it should serve as a model for future vapour cells used in near-field sensing.

We currently perform imaging with the microwave device exposed to temperatures around $140^{\circ}$C, which would be a barrier to the testing of temperature-sensitive devices. In future setups, we will move to locally heating the vapor cell with a $1.5\,\mu$m laser~\cite{Mhaskar2012}, significantly reducing the heat exposure of the device under test. If required, further reduction in operating temperature could be achieved by using LIAD techniques to modulate the Rb vapor density~\cite{Marmugi2012}.

Our microwave detection technique is not limited to $^{87}$Rb, and can be applied to any system comprised of two states coupled by a microwave transition with optical read-out of the states, including the other alkali atoms, and solid state `atom-like' systems, such as NV centres~\cite{Appel2015}.
NV center based imaging systems provide nanoscale resolution and typically work in scanning mode. They are thus complementary to our widefield imaging technique which is well adapted to image features on the micrometer scale with temporal resolution.

The full characterisation of a microwave near field requires measurements of both the electric (E$_{\mathrm{mw}}$) and magnetic (B$_{\mathrm{mw}}$) components, as there is no straightforward relationship between the components. Alkali atoms in Rydberg states have proven to be excellent sensors of E$_{\mathrm{mw}}$~\cite{Sedlacek2012,Sedlacek2013,Fan2014}, but Rydberg states are quickly destroyed in collisions with buffer gas atoms. The vapor cell requirements for B$_{\mathrm{mw}}$ and E$_{\mathrm{mw}}$ imaging would therefore seem somewhat incompatible: we require high buffer gas pressures to prevent wall relaxation and provide spatial resolution for B$_{\mathrm{mw}}$ imaging, but require that there is little to no buffer gas present for E$_{\mathrm{mw}}$ imaging. However, with the addition of a $480\,\mathrm{nm}$ laser to excite Rb Rydberg states, our control over the buffer gas inside our ultrathin cells would allow us to perform an E$_{\mathrm{mw}}$ measurement without buffer gas, then fill the cell with buffer gas and image B$_{\mathrm{mw}}$. Our setup would therefore be ideal for measurements of both components, and we would avoid the errors that using two different cell would bring, such as in cell alignment.

Microwave sensing and imaging (MSI) is an emerging field that has shown promise in a range of applications, particularly for breast cancer screening~\cite{Fear2002,Nikolova2011,Chandra2015}. Current microwave detection systems consist of an array of microwave antennas, sensitive to $E_{\mathrm{mw}}$. Optimal image reconstruction requires a high sensor density, however the density is limited by cross-talk between antennas, and by their perturbations of the microwave field. Sensor calibration is also a significant concern~\cite{Chandra2015}. Atomic sensors are not affected by any of these problems. Following the success of vapor cell magnetometers in diagnostic imaging of the heart~\cite{Bison2009,Alem2015} and brain~\cite{Savukov2013b,Johnson2013,Wyllie2012,Shah2013}, microwave imaging with vapor cells may also prove to be an attractive medical tool.

Our spatial resolution, sensitivity and distance of approach are now sufficient for characterising a range of scientific and industrial microwave devices operating at 6.8~GHz. However, frequency tunability is essential for wider applications, with industry particularly interested in imaging techniques for frequencies above 18~GHz. It is possible to use a large dc magnetic field to Zeeman shift the hyperfine ground state transitions to any desired frequency, from dc to 100s of GHz. Using a 0.8~T solenoid, we have demonstrated microwave detection up to 26.4~GHz in a proof-of-principle setup, which will be presented in a subsequent paper~\cite{Horsley2015a}.





\ack
This work was supported by the Swiss National Science Foundation (SNFS). We acknowledge helpful discussions with C.~Affolderbach, G.~Mileti, P.~Appel, M.~Ganzhorn, and P.~Maletinsky.

\section*{References}

\bibliography{bibliography_highres_Bmw}

\begin{thebibliography}{10}

\bibitem{Budker2002}
D.~Budker, W.~Gawlik, D.~F. Kimball, S.~M. Rochester, V.~V. Yashchuk, and
  A.~Weis.
\newblock {Resonant nonlinear magneto-optical effects in atoms}.
\newblock {\em Reviews of Modern Physics}, 74(October):1153, 2002.

\bibitem{Budker2007}
D~Budker and M~Romalis.
\newblock {Optical magnetometry}.
\newblock {\em Nature Physics}, 3:227--234, 2007.

\bibitem{Kitching2011}
John Kitching, Svenja Knappe, and Elizabeth~A. Donley.
\newblock {Atomic Sensors – A Review}.
\newblock {\em IEEE Sensors Journal}, 11(9):1749--1758, September 2011.

\bibitem{Kominis2003}
I~K Kominis, T~W Kornack, J~C Allred, and M~V Romalis.
\newblock {A subfemtotesla multichannel atomic magnetometer.}
\newblock {\em Nature}, 422(6932):596--9, April 2003.

\bibitem{Dang2010}
H.~B. Dang, A.~C. Maloof, and M.~V. Romalis.
\newblock {Ultrahigh sensitivity magnetic field and magnetization measurements
  with an atomic magnetometer}.
\newblock {\em Applied Physics Letters}, 97(15):151110, 2010.

\bibitem{Savukov2005b}
I.~Savukov, S.~Seltzer, M.~Romalis, and K.~Sauer.
\newblock {Tunable Atomic Magnetometer for Detection of Radio-Frequency
  Magnetic Fields}.
\newblock {\em Phys. Rev. Lett.}, 95(6):063004, August 2005.

\bibitem{Savukov2014}
I.~Savukov, T.~Karaulanov, and M.~G. Boshier.
\newblock {Ultra-sensitive high-density Rb-87 radio-frequency magnetometer}.
\newblock {\em Applied Physics Letters}, 104(2):023504, January 2014.

\bibitem{Boehi2010a}
Pascal B{\"o}hi, Max~F. Riedel, Theodor~W. H{\"a}nsch, and Philipp Treutlein.
\newblock {Imaging of microwave fields using ultracold atoms}.
\newblock {\em Appl. Phys. Lett.}, 97(5):051101, August 2010.

\bibitem{Boehi2012}
Pascal B{\"o}hi and Philipp Treutlein.
\newblock {Simple microwave field imaging technique using hot atomic vapor
  cells}.
\newblock {\em Appl. Phys. Lett.}, 101(18):181107, 2012.

\bibitem{Horsley2013}
Andrew Horsley, Guan-Xiang Du, Matthieu Pellaton, Christoph Affolderbach,
  Gaetano Mileti, and Philipp Treutlein.
\newblock {Imaging of relaxation times and microwave field strength in a
  microfabricated vapor cell}.
\newblock {\em Physical Review A}, 88(6):063407, December 2013.

\bibitem{Sedlacek2012}
Jonathon~A. Sedlacek, Arne Schwettmann, Harald K\"{u}bler, Robert L\"{o}w,
  Tilman Pfau, and James~P. Shaffer.
\newblock {Microwave electrometry with Rydberg atoms in a vapour cell using
  bright atomic resonances}.
\newblock {\em Nature Physics}, 8(11):819--824, September 2012.

\bibitem{Sedlacek2013}
J~Sedlacek, A~Schwettmann, H.~K\"{u}bler, and J.~P. Shaffer.
\newblock {Atom Based Vector Microwave Electrometry Using Rubidium Rydberg
  Atoms in a Vapor Cell}.
\newblock {\em Physical Review Letters}, 111:063001, 2013.

\bibitem{Fan2014}
H~Q Fan, S~Kumar, R~Daschner, H~K\"{u}bler, and J~P Shaffer.
\newblock {Subwavelength microwave electric-field imaging using Rydberg atoms
  inside atomic vapor cells.}
\newblock {\em Optics Letters}, 39(10):3030--3, May 2014.

\bibitem{Fear2002}
Elise~C Fear, Susan~C Hagness, Paul~M Meaney, Michal Okoniewski, and Maria~A
  Stuchly.
\newblock {Enhancing Breast Tumor Detection with Near-Field Imaging}.
\newblock {\em Microwave Magazine, IEEE}, (March):48--56, 2002.

\bibitem{Nikolova2011}
Natalia Nikolova.
\newblock {Microwave Imaging for Breast Cancer}.
\newblock {\em IEEE Microwave Magazine}, 12(7):78--94, 2011.

\bibitem{Chandra2015}
Rohit Chandra, Huiyuan Zhou, Ilangko Balasingham, Senior Member, and Ram~M
  Narayanan.
\newblock {On the Opportunities and Challenges in Microwave Medical Sensing and
  Imaging}.
\newblock {\em IEEE Transactions on Biomedical Engineering}, 62(7):1667--1682,
  2015.

\bibitem{Gajdacz2013}
Miroslav Gajdacz, Poul~L. Pedersen, Troels M{\o}rch, Andrew~J. Hilliard, Jan
  Arlt, and Jacob~F. Sherson.
\newblock {Non-destructive Faraday imaging of dynamically controlled ultracold
  atoms}.
\newblock {\em Review of Scientific Instruments}, 84(8):083105, 2013.

\bibitem{Xu2008}
Shoujun Xu, C.~Crawford, Simon Rochester, Valeriy Yashchuk, Dmitry Budker, and
  Alexander Pines.
\newblock {Submillimeter-resolution magnetic resonance imaging at the Earth’s
  magnetic field with an atomic magnetometer}.
\newblock {\em Physical Review A}, 78(1):013404, July 2008.

\bibitem{Zhao2008}
K.~F. Zhao and Z.~Wu.
\newblock {Evanescent wave magnetometers with ultrathin ($\sim 100\mu$m)
  cells}.
\newblock {\em Applied Physics Letters}, 93(10):101101, 2008.

\bibitem{Fescenko2014}
I~Fescenko and A~Weis.
\newblock {Imaging magnetic scalar potentials by laser-induced fluorescence
  from bright and dark atoms}.
\newblock {\em Journal of Physics D: Applied Physics}, 47(23):235001, 2014.

\bibitem{Hakhumyan2011}
G.~Hakhumyan, C.~Leroy, Y.~Pashayan-Leroy, D.~Sarkisyan, and M.~Auzinsh.
\newblock {High-spatial-resolution monitoring of strong magnetic field using Rb
  vapor nanometric-thin cell}.
\newblock {\em Optics Communications}, 284(16-17):4007--4012, 2011.

\bibitem{Sargsyan2014}
A.~Sargsyan, A.~Tonoyan, R.~Mirzoyan, D.~Sarkisyan, A.~M. Wojciechowski,
  A.~Stabrawa, and W.~Gawlik.
\newblock {Saturated-absorption spectroscopy revisited: atomic transitions in
  strong magnetic fields (>20 mT) with a micrometer-thin cell}.
\newblock {\em Optics Letters}, 39(8):2270, 2014.

\bibitem{Wolff2006}
Ingo Wolff.
\newblock {\em {Coplanar Microwave Integrated Circuits}}.
\newblock John Wiley $\&$ Sons, Inc., Hoboken, New Jersey, 2006.

\bibitem{Juzeliunas2006}
Eimutis Juzeliunas.
\newblock {Advances in detection of magnetic fields induced by electrochemical
  reactions—a review}.
\newblock {\em Journal of Solid State Electrochemistry}, 11(6):791--798,
  September 2006.

\bibitem{Gallo2011}
Gonzalo~E. Gallo, John~S. Popovics, and Patrick~L. Chapman.
\newblock {Corrosion monitoring of metals}.
\newblock {\em European Journal of Environmental and Civil Engineering},
  15(4):633--647, January 2011.

\bibitem{Gallo2012}
Gonzalo~E. Gallo and John~S. Popovics.
\newblock {Monitoring active corrosion of metals in natural environments with
  magnetometry}.
\newblock {\em Corrosion Science}, 63:1--4, October 2012.

\bibitem{Xu2006a}
Shoujun Xu, Simon~M. Rochester, Valeriy~V. Yashchuk, Marcus~H. Donaldson, and
  Dmitry Budker.
\newblock {Construction and applications of an atomic magnetic gradiometer
  based on nonlinear magneto-optical rotation}.
\newblock {\em Review of Scientific Instruments}, 77(8):083106, 2006.

\bibitem{Xu2006b}
Shoujun Xu, Valeriy~V Yashchuk, Marcus~H Donaldson, Simon~M Rochester, Dmitry
  Budker, and Alexander Pines.
\newblock {Magnetic resonance imaging with an optical atomic magnetometer}.
\newblock {\em Proceedings of the National Academy of Sciences of the United
  States of America}, 103(34):12668--71, August 2006.

\bibitem{Harel2008}
Elad Harel, Leif Schr\"{o}der, and Shoujun Xu.
\newblock {Novel detection schemes of nuclear magnetic resonance and magnetic
  resonance imaging: applications from analytical chemistry to molecular
  sensors.}
\newblock {\em Annual review of analytical chemistry (Palo Alto, Calif.)},
  1:133--63, January 2008.

\bibitem{Yao2010}
Li~Yao, Andrew~C Jamison, and Shoujun Xu.
\newblock {Scanning imaging of magnetic nanoparticles for quantitative
  molecular imaging.}
\newblock {\em Angewandte Chemie (International ed. in English)},
  49(41):7493--6, October 2010.

\bibitem{Yu2012}
Dindi Yu, Songtham Ruangchaithaweesuk, Li~Yao, and Shoujun Xu.
\newblock {Detecting molecules and cells labeled with magnetic particles using
  an atomic magnetometer}.
\newblock {\em Journal of Nanoparticle Research}, 14(9):1135, August 2012.

\bibitem{Yao2013}
Li~Yao, Yuhong Wang, and Shoujun Xu.
\newblock {Label-free microRNA detection based on exchange-induced remnant
  magnetization.}
\newblock {\em Chemical communications (Cambridge, England)}, 49(45):5183--5,
  June 2013.

\bibitem{Vanier1989}
Jacques Vanier, Claude Audoin, and Adam Hilger.
\newblock {\em {The Quantum Physics of Atomic Frequency Standards}}, volume~1.
\newblock Adam Hilger, Bristol, 1989.

\bibitem{Chevrollier2012}
Martine Chevrollier.
\newblock {Radiation trapping and L\'{e}vy flights in atomic vapours: an
  introductory review}.
\newblock {\em Contemporary Physics}, 53(3):227--239, May 2012.

\bibitem{Rosenberry2007}
M.~Rosenberry, J.~Reyes, D.~Tupa, and T.~Gay.
\newblock {Radiation trapping in rubidium optical pumping at low buffer-gas
  pressures}.
\newblock {\em Physical Review A}, 75(2):023401, February 2007.

\bibitem{Siddons2008}
Paul Siddons, Charles~S Adams, Chang Ge, and Ifan~G Hughes.
\newblock {Absolute absorption on rubidium D lines: comparison between theory
  and experiment}.
\newblock {\em Journal of Physics B: Atomic, Molecular and Optical Physics},
  41(15):155004, August 2008.

\bibitem{Weller2011}
Lee Weller, Robert~J Bettles, Paul Siddons, Charles~S Adams, and Ifan~G Hughes.
\newblock {Absolute absorption on the rubidium D1 line including resonant
  dipole–dipole interactions}.
\newblock {\em Journal of Physics B: Atomic, Molecular and Optical Physics},
  44(19):195006, October 2011.

\bibitem{Zentile2014}
Mark~A Zentile, Rebecca Andrews, Lee Weller, Svenja Knappe, Charles~S Adams,
  and Ifan~G Hughes.
\newblock {The hyperfine Paschen--Back Faraday effect}.
\newblock {\em Journal of Physics B: Atomic, Molecular and Optical Physics},
  47(7):075005, 2014.

\bibitem{Zelder2008}
T~Zelder, B~Geck, I~Rolfes, and H~Eul.
\newblock {Radio Science contactless vector network analysis with varying
  transmission line geometries}.
\newblock {\em Advances in Radio Science}, 6:19--25, 2008.

\bibitem{Horsley2015a}
Andrew Horsley and Philipp Treutlein.
\newblock To be published.

\bibitem{Ivanov2014}
Anton Ivanov, Thejesh Bandi, Guan-Xiang Du, Andrew Horsley, Christoph
  Affolderbach, Philipp Treutlein, Gaetano Mileti, and Anja~K. Skrivervik.
\newblock {Experimental and Numerical Studies of the Microwave Field
  Distribution in a Compact Magnetron-Type Microwave Cavity}.
\newblock {\em in proceedings of the 28th European Frequency and Time Forum
  (EFTF), Neuchatel, Switzerland, June 22-26 2014}, 2014.

\bibitem{Affolderbach2015}
Christoph Affolderbach, Guan-Xiang Du, Thejesh Bandi, Andrew Horsley, Philipp
  Treutlein, and Gaetano Mileti.
\newblock {Imaging Microwave and DC Magnetic Fields in a Vapor-Cell Rb Atomic
  Clock}.
\newblock {\em IEEE Transactions on Instrumentation and Measurement}, PP(99),
  2015.

\bibitem{Horsley2015thesis}
Andrew Horsley.
\newblock {High Resolution Field Imaging with Atomic Vapor Cells}.
\newblock {\em PhD Thesis, Department of Physics, University of Basel,
  Switzerland}, 2015.

\bibitem{Horsley2013a}
Andrew Horsley, Guan-Xiang Du, Matthieu Pellaton, Christoph Affolderbach,
  Gaetano Mileti, and Philipp Treutlein.
\newblock {Spatially Resolved Measurement of Relaxation Times in a
  Microfabricated Vapor Cell}.
\newblock {\em Proceedings of the 2013 Joint IEEE-UFFC, EFTF and PFM
  Symposium}, pages 575--578, 2013.

\bibitem{Chrapkiewicz2014}
Radoslaw Chrapkiewicz, Wojciech Wasilewski, and Czeslaw Radzewicz.
\newblock {How to measure diffusional decoherence in multimode Rubidum vapor
  memories?}
\newblock {\em Optics Communications}, 317:1--6, 2014.

\bibitem{Ishikawa2000}
Kiyoshi Ishikawa and Tsutomu Yabuzaki.
\newblock {Diffusion coefficient and sublevel coherence of Rb atoms in N$_2$
  buffer gas}.
\newblock {\em Phys. Rev. A}, 62(6):065401, November 2000.

\bibitem{Mikhailov2009}
Eugeniy~E Mikhailov, I~Novikova, M~D Havey, and F~A Narducci.
\newblock {Magnetic field imaging with atomic Rb vapor.}
\newblock {\em Opt. Lett.}, 34(22):3529--31, November 2009.

\bibitem{Allred2002}
J.~Allred, R.~Lyman, T.~Kornack, and M.~Romalis.
\newblock {High-Sensitivity Atomic Magnetometer Unaffected by Spin-Exchange
  Relaxation}.
\newblock {\em Physical Review Letters}, 89(13):130801, September 2002.

\bibitem{Mhaskar2012}
R.~Mhaskar, S.~Knappe, and J.~Kitching.
\newblock {A low-power, high-sensitivity micromachined optical magnetometer}.
\newblock {\em Applied Physics Letters}, 101(24):5--8, 2012.

\bibitem{Marmugi2012}
Luca Marmugi, Silvia Gozzini, Alessandro Lucchesini, Andrea Bogi, Alessia
  Burchianti, and Carmela Marinelli.
\newblock {All-optical vapor density control for electromagnetically induced
  transparency}.
\newblock {\em Journal of the Optical Society of America B}, 29(10):2729,
  September 2012.

\bibitem{Appel2015}
Patrick Appel, Marc Ganzhorn, Elke Neu, and Patrick Maletinsky.
\newblock {Nanoscale microwave imaging with a single electron spin in diamond}.
\newblock {\em to be published}.

\bibitem{Bison2009}
G.~Bison, N.~Castagna, A.~Hofer, P.~Knowles, J.-L. Schenker, M.~Kasprzak,
  H.~Saudan, and A.~Weis.
\newblock {A room temperature 19-channel magnetic field mapping device for
  cardiac signals}.
\newblock {\em Applied Physics Letters}, 95(17):173701, 2009.

\bibitem{Alem2015}
Orang Alem, Tilmann~H Sander, Rahul Mhaskar, John LeBlanc, Hari Eswaran, Uwe
  Steinhoff, Yoshio Okada, John Kitching, Lutz Trahms, and Svenja Knappe.
\newblock {Fetal magnetocardiography measurements with an array of
  microfabricated optically pumped magnetometers}.
\newblock {\em Physics in Medicine and Biology}, 60(12):4797--4811, 2015.

\bibitem{Savukov2013b}
I~Savukov and T~Karaulanov.
\newblock {Magnetic-resonance imaging of the human brain with an atomic
  magnetometer.}
\newblock {\em Applied Physics Letters}, 103(4):43703, July 2013.

\bibitem{Johnson2013}
Cort~N Johnson, P~D~D Schwindt, and M~Weisend.
\newblock {Multi-sensor magnetoencephalography with atomic magnetometers.}
\newblock {\em Physics in medicine and biology}, 58(17):6065--77, September
  2013.

\bibitem{Wyllie2012}
Robert Wyllie, Matthew Kauer, Ronald~T Wakai, and Thad~G Walker.
\newblock {Optical magnetometer array for fetal magnetocardiography.}
\newblock {\em Optics letters}, 37(12):2247--9, July 2012.

\bibitem{Shah2013}
Vishal~K Shah and Ronald~T Wakai.
\newblock {A compact, high performance atomic magnetometer for biomedical
  applications.}
\newblock {\em Physics in medicine and biology}, 58(22):8153--61, November
  2013.

\end{thebibliography}

\end{document}